\documentstyle[12pt,psfig,twoside]{article}
\newcounter{muni}
 
\begin{document}\hbadness=10000\pagenumbering{arabic}
\pagestyle{myheadings}\markboth{J. Letessier, J. Rafelski 
and A. Tounsi}{Strange Particle Freeze-Out at AGS}
\title{\vspace*{-1cm}
STRANGENESS AND PARTICLE FREEZE-OUT\\
IN NUCLEAR COLLISIONS AT 14.6 GeV A}
\author{$\ $\\
{\bf J.  Letessier}$^1$, {\bf J. Rafelski}$^{1,2}$ {\rm and} 
{\bf A. Tounsi}$^1$\\ $\ $\\
$^1$Laboratoire de Physique Th\'eorique et Hautes Energies\thanks{\em
Unit\'e  associ\'ee au CNRS UA 280, \rm\newline \hspace*{0.5cm}
Postal address: LPTHE~Universit\'e PARIS 7, Tour 24, 5\`e
\'et., 2 Place Jussieu, F-75251 CEDEX 05.}~, Paris\\
$^2$Department of Physics, University of Arizona, Tucson, AZ 85721\\}
\date{}   
\maketitle 
\vspace{-9cm}\noindent {PAR/LPTHE/94--15}  
\hfill Published in Phys. Lett. B328 (1994) 499\vspace{8cm} 
\begin{abstract}
\noindent{
We study the chemical conditions at freeze-out associated with the
production of strange-particles in Si--Au collisions at 14.6 GeV A.
We obtain freeze-out chemical potentials and 
temperature, and determine the entropy as well as the final particle 
abundance. We also consider in detail the 
alternative evolution scenarios involving the hadronic gas and
the deconfined phase. 
}\end{abstract}  
{\bf 1. Introduction.} In the past years a number of experiments at
14.6~GeV~A (AGS-BNL) and at 200 GeV A (SPS-CERN) have shown that the
strange particle abundance differs considerably in nuclear interactions
from that in nucleon-nucleon collisions \cite{Quer93}.  Qualitative
agreement with the experiment within the ARC particle cascade models has
been reached at the AGS energy \cite{ARC,Gon94}, without need for any novel
mechanisms of strange particle production enhancement, but at CERN
energies such agreement has not been obtained \cite{GazP} 
--- moreover so far  no
microscopic model has emerged, capable to describe in all necessary
detail the multi-strange anti-baryon abundance seen at CERN in the
central rapidity region. Furthermore one finds significant 
difference in the strangeness flow comparing 14.6 GeV A with 200 GeV A
collisions \cite{RD93}.

The ARC results indicate transient presence of a clearly defined 
region of high density matter at central rapidity, seen in
particular in the case of the Au--Au collision simulations.
This suggests that a local equilibrium thermal model
(fireball model) may also do well in
describing these collisions and in this work we will assume simply that
all central rapidity strange particles originate from a central, thermally
equilibrated fireball --- our analysis will allow in principle for chemical 
strange particle abundance deviations from full equilibrium here assumed. 
We implemented here several features which are more difficult to account 
for in the full microscopic simulation:
\begin{itemize}

\vspace{-0.3cm}
\item {we incorporate the quantum gas nature for the hadronic
particles;}

\vspace{-0.3cm}
\item {we treat the entropy contained in the initial thermal
state as a parameter and determine it as needed to describe the
data;}

\vspace{-0.3cm}
\item {we explore the consistency of the description with 
the possible transient presence of a novel form of matter, such as is the 
locally deconfined quark-gluon plasma (QGP).}
\end{itemize}

The parameters of the thermal model are the temperature $T$ and the
chemical potentials $\mu_i$ of the different quark flavors. In this
discussion it is not necessary to distinguish the $u$ and $d$ flavors and
we therefore consider the light quark chemical potential $\mu_{\rm q}$
and the related fugacity $\lambda_{\rm q}=\exp(\mu_{\rm q}/T)$. Since
three quarks together have baryon number one, the baryon number fugacity
is $\lambda_{\rm B}=\lambda_{\rm q}^3$, that is $\mu_{\rm B}=3\mu_{\rm
q}$.  A finite value of $\mu_{\rm B}$ indicates that we have a finite
baryon density. 
 
We will study in some detail strange particles which contain
strange quarks (or anti-quarks),  associated with strange quark
fugacity and chemical potential: $\lambda_{\rm s}=\exp(\mu_{\rm s}/T)$.
At relative particle abundance equilibrium the
value of $\mu_{\rm s}$ is fixed within our approach 
by the requirement that strangeness is conserved. 
In any hadronic-gas (HG) picture of the reaction leading to high
baryon density, there exists typically a considerable asymmetry in the
strange and anti-strange particle phase space and hence the requirement
of strangeness conservation normally leads to a considerable value of
$\mu_{\rm s}^{\rm HG}>0$.  An exception from this rule exists and is
discussed in detail elsewhere \cite{LTR,Cley92}. On the other hand, since
in the deconfined QGP state both $s$ and $\bar s$ quarks are free,
one always has $\mu_{\rm s}^{\rm QGP}=0$, assuming that strangeness is
conserved, $<s-\bar s>=0$, on the time scale of A--A collisions. 
Said differently, observation of $\mu_{\rm s}\ne 0$ (that is strange-quark fugacity
$\lambda_{\rm s}\ne 1$) strongly suggests that these particles emerge from
a final state which is a confined HG phase.
 
In the analysis of the collisions of S--ions at 200 GeV A with different
nuclear targets carried out at CERN-SPS one finds in the framework of the
thermal model \cite{Raf91,27R} that the strange-quark fugacity is
$\lambda_{\rm s}\simeq 1$ (or said alternatively, the strange-quark
chemical potential $\mu_{\rm s}\simeq 0$). Furthermore, considering
particle production one finds that these collisions are leading to very
entropy rich state \cite{Let93l}. Finally, also the
strange-particle phase space occupancy is close to equilibrium, 
($\gamma_{\rm s}\simeq 1$) \cite{Raf91,27R} 
--- this suggests the presence of an unusually effective
strangeness production mechanisms expected only in the QGP phase. These
three results strongly suggest that QGP phase is formed in collisions at
200 GeV A, but considerable additional effort is needed to show this to
be the case. Importantly, $\mu_{\rm s}=0$ is found, despite the
expectation that in (a slow) hadronization process of the QGP the memory
about the transient, deconfined state would be erased and hadronic
particle abundance would reflect conditions of an equilibrated HG. Thus
the value $\mu_{\rm s}=0$ also suggests (if we presume 
formation of QGP like state) that at CERN the hadronization is
sudden and that therefore certain hadronic observables could carry
considerable information about the primordial phase.  One of our principal 
results here is a demonstration that as the energy decreases from 
$\sqrt{s_{\rm NN}}=8.8$ GeV, at CERN,
to $\sqrt{s_{\rm NN}}=2.5$ GeV, at AGS,
this situation changes with the freeze-out well described by equilibrium HG.
This may be considered to indicate that at AGS:

\begin{itemize}

\vspace{-0.3cm}
\item {the QGP phase was not formed at all, or,}
 
\vspace{-0.3cm}
\item {that complete re-equilibration 
occurs when the primordial QGP hadronizes.}
 
\vspace{-0.3cm}
\end{itemize}
 
While the freeze-out chemical potentials at CERN energies
are well known, the freeze-out temperature remains yet to be determined,
despite the relatively large sample of data:  the results
indicate that the hadronization of the primordial phase is sudden, and
one has to permit non-equilibrium abundances of diverse hadronic
particles. This introduces additional off-equilibrium parameters which
need to be determined along with the freeze-out temperature \cite{LTR}.
On the other hand there seems to be no need to introduce 
non-equilibrium particle abundance parameters (other than strange phase
space occupancy factor $\gamma_{\rm s}$) at AGS  energies and hence 
we can determine in this work the freeze-out temperature applicable to
AGS --- we note that the inverse slope parameters of particle spectra,
interpreted within the fireball model are the blue shifted 
freeze-out temperatures $T_{\rm f}$, the Doppler shift arising from the 
collective flow phenomena~\cite{SH92}.
 
In this work we will employ state of the art
equations of state for both HG and QGP fireballs:
 
\begin{itemize}

\vspace{-0.3cm}
\item {for HG we use a refined Hagedorn approach 
which accounts for
interactions by treating all hadronic resonances as independent
particles. We include numerically the known resonances, considered
usually to be Boltzmann particles \cite{LTR}, but in the present work
which reaches to lower temperatures $T\le m_\pi$, where $m_\pi$ is the
pion mass, pions and all mesons are treated as bosons and we also 
treat nucleons and $\Delta$-resonances as fermions.} 

\vspace{-0.3cm}
\item {for QGP we take the analytical form of an ideal quark-gluon gas, 
but amend the number of degrees of freedom to reflect
the reduction due to perturbative QCD-interactions, and we 
introduce non-perturbative thermal mass for quarks and  gluons 
\cite{Let94b}.}
\end{itemize}
 
\noindent{\bf 2. AGS data}. We will discuss here Si--Au results
obtained at 14.6 GeV A by the experiment E802/E859 at BNL, 
which were presented at the QM'93
meeting \cite{newdata1,newdata2}, in particular:

\begin{itemize}

\vspace{-0.3cm}
\item {the ratio K$^-$/K$^+ = 0.22 \pm 0.02$, 
which is constant in the
central rapidity range $1<y<1.6$, consistent with the hypothesis that 
the dominant particle source is a fireball. 
This ratio varies strongly in target and
projectile rapidity regions, as can be expected due to secondary
production mechanisms and the presence of baryon-rich reaction
fragments;}

\vspace{-0.3cm}
\item {the central rapidity ratio
$\overline{\Lambda}/\Lambda=(2.0\pm0.8) 10^{-3}$
averaged over the acceptance of the spectrometer ($1.1<y<1.7$).}
\end{itemize}
 
We will now use these strange particle ratios to determine the values of
chemical potentials at freeze-out; we have \cite{LTR,Cley92,Raf91,27R}:
\begin{eqnarray}
{{\overline \Lambda}\over{\Lambda}}\hspace{-0.6cm}&&= 
\lambda_{\rm s}^{-2}\lambda_{\rm q}^{-4}\,,\label{lamrat}\\
&~&\nonumber\\
{{\rm K^-}\over {\rm K^+}} \hspace{-0.6cm}&&
= \lambda_{\rm s}^{2} \lambda_{\rm q}^{-2}\,,\label{krat}
\end{eqnarray}
We obtain \cite{RD93}:
\begin{eqnarray}
\lambda_{\rm s} \hspace{-0.6cm}&&
={{\left({\rm K}^-/{\rm K}^+\right)^{1/3}}\over 
{\left(\overline{\Lambda}/\Lambda\right)^{1/6}}}
=1.71\left\{
\begin{array}{rl}
+0.20&\ \\-0.15&\ 
\end{array}\right. \,,\label{numlams}\\
&~&\nonumber\\
\lambda_{\rm q}\hspace{-0.6cm}&&
={{\left({\rm K}^+/{\rm K}^-\right)^{1/6}}\over 
{\left(\overline{\Lambda}/\Lambda\right)^{1/6}}}
=3.62\left\{
\begin{array}{rl}
+0.39&\ \\-0.26&\ 
\end{array}\right. \,. \label{numlamq}
\end{eqnarray}
These two results are considerably different from the result obtained at
CERN energies: $\lambda_{\rm s}=1.03\pm 0.05,\, \lambda_{\rm q}=1.49\pm 0.05$. 
The AGS result for $\lambda_{\rm s}$ differs by more than four standard 
deviations from the CERN value --- we note that the 
high-power roots entering above analysis reduce the significance of the
large errors in the data. We further note the relatively high value of
$\lambda_{\rm q}$ which is more than twice as large as that of the CERN
experiments and suggestive of considerable baryon density reached at AGS
conditions.  

An interesting implication and to some extent a verification 
of the thermal fireball model is that the two particle ratios
we consider directly determine the $\bar p/p$ ratio --- note that
Eqs.~(\ref{lamrat}),~(\ref{krat}) imply:
\begin{eqnarray}
{\overline{p}\over p}=\lambda_{\rm q}^{-6} =
{\overline{\Lambda} \over \Lambda} {{\rm K}^-\over {\rm K}^+} 
=(0.44\pm0.2)10^{-3}\,.
\end{eqnarray}
We note \cite{newdata1} that at central rapidity $y=1.3$ the 
$\bar p$-yield is given to be  $dN/dy\sim 10^{-2}$,
which, when combined with the central proton yield $dN/dy\sim 20$,
leads to $\bar p/p=(0.5\pm0.2)~10^{-3}$
(where we estimate the error from the results shown in \cite{newdata1}), 
which agrees in a remarkable way 
with the result we obtained above. Since  within the error bar the ratio 
varies by factor two, it is also consistent with the other experimental
results reported \cite{ap814}. We stress that this agreement of the fireball 
model with experiment may be 
fortitouse, since the $\bar p$ data are still preliminary, while the 
$p$-yield is strongly influenced by the target spectator tail. We 
will therefore return to a full discussion of the validity of the 
fireball model when the Au--Au results become available.\\
 
\noindent{\bf 3. Theoretical interpretation.} A natural question arising
here is in how far these values of $\lambda_{\rm s},\ \lambda_{\rm q}$
are consistent with the expected properties of the hadronic state and
what they tell us about other freeze-out conditions. First, let us note a
powerful constraints for our analysis: the three thermal parameters $T,\
\lambda_{\rm s},\ \lambda_{\rm q}$ are constrained by the requirement
that the net strangeness of the source be zero --- roughly speaking, the
sum of hyperons and anti-kaons must equal the number of kaons, (we of course
account for all other strange particles) and the values found in
Eqs. (\ref{numlams}), (\ref{numlamq}) are consistent with this
constraint. When implementing in detail the strangeness conservation
constraint we allow for one off-equilibrium parameter: since
strange quark flavor is produced relatively slowly in hadronic
interactions (but strangeness produced is easily redistributed
among different degrees of freedom), it is customary \cite{Raf91}
to weight with a factor $\gamma_{\rm s}^n$ all hadrons containing $n$ 
strange or anti-strange quarks. This strangeness phase space occupancy 
factor enters in some theoretical results \cite{LTR},thus 
we will allow in the discussion below a range of values: 
on the low side we take $\gamma_{\rm s}=0.2$, which is the value
typical of $p$-$p$ collisions, and the upper limit will be 
$\gamma_{\rm s}=0.7$, which is found in 200 GeV A experiments. 
The equilibrium value is $\gamma_{\rm s}^{\rm eq}=1$. 
 
A parameter of all thermal models is the specific entropy (entropy per
baryon) content of the source: once the initial thermalized state is
established, its evolution to the freeze-out condition is proceeding
nearly without entropy production as we have found studying the approach
to chemical equilibrium for (isolated) fireballs \cite{Cool}.
A most remarkable property of the HG (and also QGP) equations of state 
of hadronic matter which we use here is that at a given value of specific 
entropy per baryon (given strangeness conservation constraint)  
corresponds to a nearly constant ratio:
\begin{equation}
R_{\lambda}=\lambda_{\rm q}/\lambda_{\rm s}\,.
\end{equation}
In the QGP phase this is well understood \cite{Raf91,Let94b} --- a 
dimensionless quantity, such as is $S/B$ can only be a function of 
dimensionless objects, and since $\lambda_{\rm s}=1$, it is only a 
function of $\lambda_{\rm q}$. In HG this finding which implies
a large degree of cancellation, was for us totally unexpected. In Fig.
\ref{F1} we show for each value of specific entropy $10\le S/B\le 30$ the
range of $R_\lambda=\lambda_{\rm q}/\lambda_{\rm s}$ which is possible
when we vary temperature in the wide range $T\in (100,\,210)$ MeV and
also allow for a wide range in strange particle occupancy $\gamma_{\rm
s}\in (0.2,\,0.7)$ (lower bars are for $\gamma_{\rm s}=0.2$ and upper 
ones for 0.7). We have interpreted in Eq.(\ref{krat}) the kaon ratio
in terms of fugacities. This implies $R_\lambda=2.12\pm0.12$
and we show (right two vertical lines in Fig.\,\ref{F1}) 
that this corresponds to the specific entropy at freeze-out 
$S/B=13\pm 1$ (for $\gamma_{\rm s}\sim 0.4$). 
This implies after all resonances have decayed a final state  ratio 
of mesons to baryons  $R_{\rm m/b}=1.6\pm0.2\,$,
which is probably compatible with the somewhat lower value 1.2
reported \cite{Gon94}; since in our calculus we account for the 
enhancement of low $p_\bot$ pions, arising from resonance disintegration. 
In Fig. \ref{F1} we also show the value of $R_\lambda=
\lambda_{\rm q}/\lambda_{\rm s}\simeq 1.5$ associated with the 200 GeV A
results (left two vertical lines in Fig.\,\ref{F1})
--- we see that the HG type interpretation at CERN would imply 
a specific entropy content of $24\pm 5$, while the high particle flow seen 
indeed points towards $S/B=40$ more appropriate for 
a rapidly disintegrating QGP phase \cite{Let93l}.
  
In order to determine the freeze-out temperature $T_{\rm f}$, we consider 
(see Fig. \ref{F2}) the ratio $\overline{\Lambda}/\Lambda$ at fixed $S/B$ 
as function of $T$. Again we find that there is residual
$\gamma_{\rm s}$ dependence --- the thick lines are
for the upper limit $\gamma_{\rm s}=0.7$, the thin lines for the
lower limit $\gamma_{\rm s}=0.2$.  The short dashed (upper) lines are for 
$S/B=14$, the long dashed (lower) ones for $S/B=12$ and the
solid lines are for $S/B=13$. The experimental
range  $\overline{\Lambda}/\Lambda\in (1.2,\,2.8)\,10^{-3}$ thus leads to
the range $T_{\rm f}=127\pm8$ MeV, and we indicate the
experimental and theoretical ranges of the values by the error 
cross in Fig. \ref{F2}. Given the thermal freeze-out conditions we 
can obtain the freeze-out baryon density $\rho_{\rm f}=(0.7\pm0.3)\rho_N$,
where $\rho_N=0.16\,{\rm fm}^{-3}$.
 
We now turn to the two questions:
 
1) what are the initial thermal conditions when the fireball is formed? 

2) what differences, if any, are there between HG and QGP phases?

In Fig. \ref{F3} we show in the $\mu_{\rm B},\, T$ plane for the HG case
(thick lines) and QGP case (thin lines) the different possible histories of
the collision. The dashed lines correspond to a fixed energy per baryon 
2.55 GeV A in the CM frame, while the solid lines are for fixed specific 
entropy per baryon $S/B=13$. Where the solid and dashed lines meet (within a 
trapezoidal region determined by one unit error in entropy and an
error of 0.15 GeV in CM energy) we have a consistency between initial 
energy and final state entropy, thus presumably these are the initial values
of thermal parameters. The initial HG
state (for $S/B=13,\, E/B=2.55$ GeV) has an energy density 
$\varepsilon_0^{\rm HG}=2.3$ GeV fm$^{-3}$, and baryon density
$\rho_0^{\rm HG}=6\rho_N$.
The initial pressure is $P_0^{\rm HG}=0.3$ GeV fm$^{-3}$, which is normally
considered a value typical of the deconfined phase. In the QGP phase we find 
accordingly: $\varepsilon_0^{\rm QGP}=1.2$ GeV fm$^{-3}$, 
$\rho_0^{\rm QGP}=2.9\rho_N$, $P_0^{\rm QGP}$=0.39 GeV fm$^{-3}$.
Surprizingly, the QGP is the more dilute phase at these conditions,
but it can be argued that the greater stiffness of QGP, as compared to 
HG, reduced the nuclear compression reached in collision.

Remarkably, we find a very large 
difference in $\mu_{\rm B}$ which takes an initial value  
$\mu_0^{\rm HG}=440\pm40$~MeV 
in the HG scenario and $\mu_0^{\rm QGP}=910\pm150$~MeV 
in the QGP case --- not shown in the
Fig. \ref{F3} is that the strangeness conservation requirement 
leads in the HG to an {\it initial} value 
$\mu_{\rm s,0}^{\rm HG}=0$, just as is in the case of the QGP. 
We see in Fig. \ref{F3} that the initial temperatures $T_0$ for QGP
and HG scenarios are practically equal. For $E/B=2.55\pm0.15$ GeV 
and $S/B=13\pm1$ we have $T_0=190\pm30$ MeV.
We recall that the effective temperature seen in the $m_\bot$ particle 
spectra is nearly $T_0$ \cite{27R,SH92}.
 Thus presence of
high inverse slopes in high $m_\bot$ spectra with $T\simeq
225$ MeV \cite{newdata1} 
is consistent with our considerations and both equations of 
state used (HG and QGP). Perhaps a small edge can be given to the QGP case,
since one should allow in principle for some entropy production in the 
transition to HG and for $S/B=11$ one would indeed reach with the central
value  to $T_0^{\rm QGP}\simeq 215$ MeV.

The QGP fireball at $S/B=13$ (thin solid line) evolves  practically at fixed 
$\lambda_{\rm q}=4.8$ and $\lambda_{\rm s}=1$. For the HG fireball at fixed 
$S/B=13$ (thick solid line) there is a strong variation in both these 
fugacities but as noted in Fig. \ref{F2} the ratio 
$R_\lambda^{\rm HG}(=\lambda_{\rm q}/\lambda_{\rm s})=2.17$ remains 
practically constant. 
The `experimental' cross is set at $\mu_{\rm B,f}=485\pm 70$~MeV 
and $T_{\rm f}=127\pm8$~MeV corresponding to our determination of freeze-out 
conditions (with $\mu_{\rm s,f}=68$ MeV). This appears near the region
where the two evolution curves meet (QGP and HG) but in fact both miss each
other greatly in a three dimensional display, since the ratio 
$R_\lambda^{\rm QGP}=\lambda_{\rm q}/\lambda_{\rm s}=4.8$ 
on the QGP path-line and $=2.17$ on the HG path-line. 
 
We have also studied where a possible phase transformation from 
QGP to HG may be if QGP is produced. 
Since we are in relative low pressure domain, the pressure
of the QGP phase is impacted by the vacuum {\it bag constant 
${\cal B}$} contribution. We took 
${\cal B}=0.1\ {\rm GeV\,fm}^{-3}$ (corresponding to 
${\cal B}^{1/4}=170$ MeV). From the resulting phase diagram we 
obtained that in the event that a first order phase transition 
is occurring, it would be at $P=0.04\pm0.01\ {\rm GeV\,fm}^{-3}$ 
and this determines points on
the two evolution paths shown, connected by nearly vertical lines --- we note
that there is minor re-heating occurring while the baryo-chemical potential 
drops by 15\% --- however, the major re-equilibration is in the jump from 
$\lambda_{\rm s}=1$ in the plasma to $\lambda_{\rm s}\sim1.7$ in the HG phase
(as discussed above this is still amplified by comparing the greatly 
different  values of $R_\lambda$ in both phases). 
Thus the `short' connection between the
QGP to HG paths would be considerably stretched in full three dimensional 
display, reflecting on the need to well re-equilibrate the matter 
in transition. Therefore we believe that
there is not a slightest remembrance of 
the possible QGP history in the AGS data considered here. 

However, it is likely that the presence of a primordial 
phase can be in principle seen in hadronic  particle spectra due to the great 
difference in the $R_\lambda$ ratio  between QGP and HG cases --- 
the difference is by factor 2.2. Thus e.g. the ratio  of K$^+$ to
$\pi^+$ particle abundance at fixed $m_\bot$ should increase at large
$m_\bot$  supposing that the high $m_\bot$ particles are directly emitted 
from the possibly formed QGP phase. Since in the fireball model
\begin{equation}
\left. {{\rm K}^+\over \pi^+}\right\vert_{m_\bot}=R_\lambda
\end{equation}
we should see that the high $m_\bot$ ratio decreases just by factor 2.2 
as the high $m_\bot$ QGP primordial contribution are overcome by 
hadronization yields, and at still smaller $m_\bot$
by even more as the pion abundance is diluted by the $\Delta$ 
decays (an effect which we can fully account for given the model for 
the freeze-out conditions we developed).\\

\noindent{\bf 4. Conclusions and outlook.} We have reached a full and 
satisfactory description of the freeze-out conditions at AGS energies, 
employing thermal fireball model. 
We have alluded to a number of consistency checks, e.g. have mentioned 
that the meson to baryon ratio in the final state 
$R_{\rm m/b}=1.6\pm 0.2$, and also noted that the  ratio $\bar p/p=0.5\,10^{-3}$
is of the correct magnitude.  We note that given the
freeze-out conditions we can compute the abundances of most hadronic
particles in the final state, and we intend to develop a more 
detailed and comprehensive analysis once more
precise input data  become available. 
 
Should QGP be formed at these low energies, 
we hope that akin to the situation at 200~GeV~A, one
will be able to see considerable anomalies in abundances of multi-strange
anti-baryons. These are produced extremely rarely in the HG environment
and normally one would not believe that they can be observed for beam
energies at about 14.6 GeV A. However, should at AGS indeed the QGP
phase be formed, their abundance would be only little
reduced as compared to the 200 GeV~A situation, while the experimental
sensitivity would be greatly amplified, since the entropy content and
hence the final state multiplicity is much smaller at these lower
energies. Given the theoretical results we have presented regarding the
initial thermal conditions reached in the central fireball in 14.6 GeV~A
collisions, it is hard not to envision QGP formation at AGS and hence we
would like to encourage strongly that a search be made for these hadronic
observables \cite{antibaryons} of the QGP phase. Thus
we are in particular interested in
central rapidity data about (multi-)strange anti-baryons which we hope
could help resolve the question if there has been formation of transient
deconfined phase. In particular we may be able to determine the value of
$\gamma_{\rm s}$; we need to look at the ratios: 
\begin{eqnarray} 
R_{\bar {\rm s}}=\gamma_{\rm s}{\lambda_{\rm q}\over \lambda_{\rm s}} = 
\gamma_{\rm s}R_\lambda
\sim \left.\frac{1}{2}\,{\overline{\Lambda}+\overline{\Sigma^0}
\over\overline{p}}\right\vert_{m_\bot}
\sim \left.2\,{\overline{\Xi^-}\over\overline{\Lambda}+\overline{\Sigma^0} } 
\right\vert_{m_\bot}
\sim \left.\frac{1}{2}\,{\overline{\Omega^-}\over \overline{\Xi^-}} 
\right\vert_{m_\bot}
\,. 
\end{eqnarray}
Resonance decays influence these yields, and hence the
above relation holds approximately, but increasingly exactly as the
strangeness content increases.  If QGP is formed and the strangeness
phase space were fully saturated, we would expect that the $R_{\rm \bar s}$
ratio is indeed greater than unity. 
Due to the large value of $\lambda_{\rm q}$ we found, if QGP is formed with 
$\gamma_{\rm s}\simeq 0.7$ and $R_\lambda=\lambda_{\rm q}/\lambda_{\rm s}
=4.8$ governing emission at high $m_\bot$, we obtain $R_{\rm\bar s}\simeq3$.
Using the HG values $\gamma_{\rm s}\sim 0.15$, $R_\lambda=2.1$ 
one finds a yield typical for  $p$-$p$
reactions, with $R_{\bar{\rm s}}\sim 0.3$. 
We recall here that for the CERN experiment
the corresponding discussion addressed the ratio of yields of
$\overline{\Xi}/\overline{\Lambda}$ and the high value of $\gamma_{\rm
s}\sim 0.7 $ found there arises from the remarkable result that 
$R_{\bar{\rm s}}$ is ${\cal O}(1)$ at 200 GeV.
 
Even though we have here discussed only the Si beam experiments 
at 14.6 GeV~A involving collisions with the heaviest target (Au) as is
appropriate for the here considered fireball model,
our analysis can be easily continued to the case of 
Au--Au collisions at 11--12 GeV~A \cite{Gon94,newdata3}. Because of the
projectile--target symmetry the CM specific energy (assuming full
stopping or equal baryon and energy stopping) is for the generally
considered value of 11.7 GeV A exactly the
value we assumed here, namely $\sqrt{s_{\rm NN}}=2.55$~GeV. We note that a
slight increase in the ratio of
${\rm K}^+/{\rm K}^-$ from 4.5 to perhaps 5 would imply a slight drop in
the specific entropy (see Eq.\,(\ref{krat}) and Fig. \ref{F1}), 
perhaps by 1/2 unit. This
causes only incremental changes in all freeze-out results and thus there
will be little change in final state particle abundances.
However, we find, within the framework of HG model, 
a considerable sensitivity of the initial conditions 
(baryon density, pressure and energy density), seen as function of the 
entropy content: 
a drop by one unit of $S/B$ from 13 to 12 implies doubling of all these
initial values. Thus it is well possible
that much more extreme conditions are reached in the Au--Au collisions as
compared to Si--Au, even if the reaction products and multiplicity
appear to be similar for both systems.

In summary: we have used that results on
strange particle ratios ${\overline \Lambda}/{\Lambda}$ and ${\rm
K^-}/{\rm K^+}$ from AGS in a fireball model framework and have
determined freeze-out values of temperature  ($T_{\rm f}=127\pm8$ MeV), 
and chemical potentials ($\mu_{\rm B,f}=485\pm70$ MeV and 
$\mu_{\rm s,f}=68(+20,-16)$ MeV, final state specific entropy $S/B=13\pm1$ 
accompanied by meson to baryon  particle
ratio $R_{\rm m/b}=1.6\pm0.2$, and also within the HG and QGP models, the 
(extreme) initial conditions prevailing when the fireball is formed,
in particular $T_0=190\pm 30$~MeV.\\
 
\noindent{\bf Acknowledgements.}  Work of JR is in part supported by 
US-Department of Energy under grant DE-FG02-92ER40733. JR thanks his 
colleagues for their kind hospitality in Paris.
 

\begin{figure}[h]
\vspace*{2cm}
\centerline{\bf FIGURE CAPTIONS}
\caption[SB-HG]
{
\label{F1}
For a given specific entropy per baryon $S/B$ we show the range of
allowable values $\lambda_{\rm q}/\lambda_{\rm s}$ values in HG phase,
when temperature is varied between 100 and 210 MeV and the strangeness
occupancy factor $\gamma_{\rm s}$ between is 0.2 (lower values) and 0.7
 (upper values). Strangeness conservation constraint is imposed. 
See text for discussion of vertical lines.
}
\caption[Tf] 
{
\label{F2}
The ratio $\overline{\Lambda}/\Lambda$  as function of temperature, at
fixed specific entropy and for $\gamma_{\rm s}=0.7$ (thick lines) 
and 0.2 (thin lines). Solid lines $S/B=13$, short dashed 
$S/B=14$, long dashed $S/B=12$. 
}
\caption[Tmu-HG-QGP] 
{
\label{F3}
$\mu_{\rm B}$--$T$ plane with thick  lines: HG model; thin lines: QGP model. 
Dashed lines: fixed energy per baryon $E/B=2.55$ GeV, upper for HG, 
lower for QGP.
Solid lines: fixed specific entropy $S/B=13$, upper for QGP, lower for 
HG model.
Trapezoidal regions enclose the initial condition (see text). Solid line 
connecting
QGP and HG with dotted lines left/right: possible phase transition at 
$P=0.04\pm0.01$ GeV/fm$^3$. Note that continuations of $S/B=13$ lines 
beyond transition or hadronization points are dotted.
}

\vfill
\end{figure}
\vfill\eject

\begin{figure}[p]
\centerline{\hspace{0.2cm}\psfig{figure=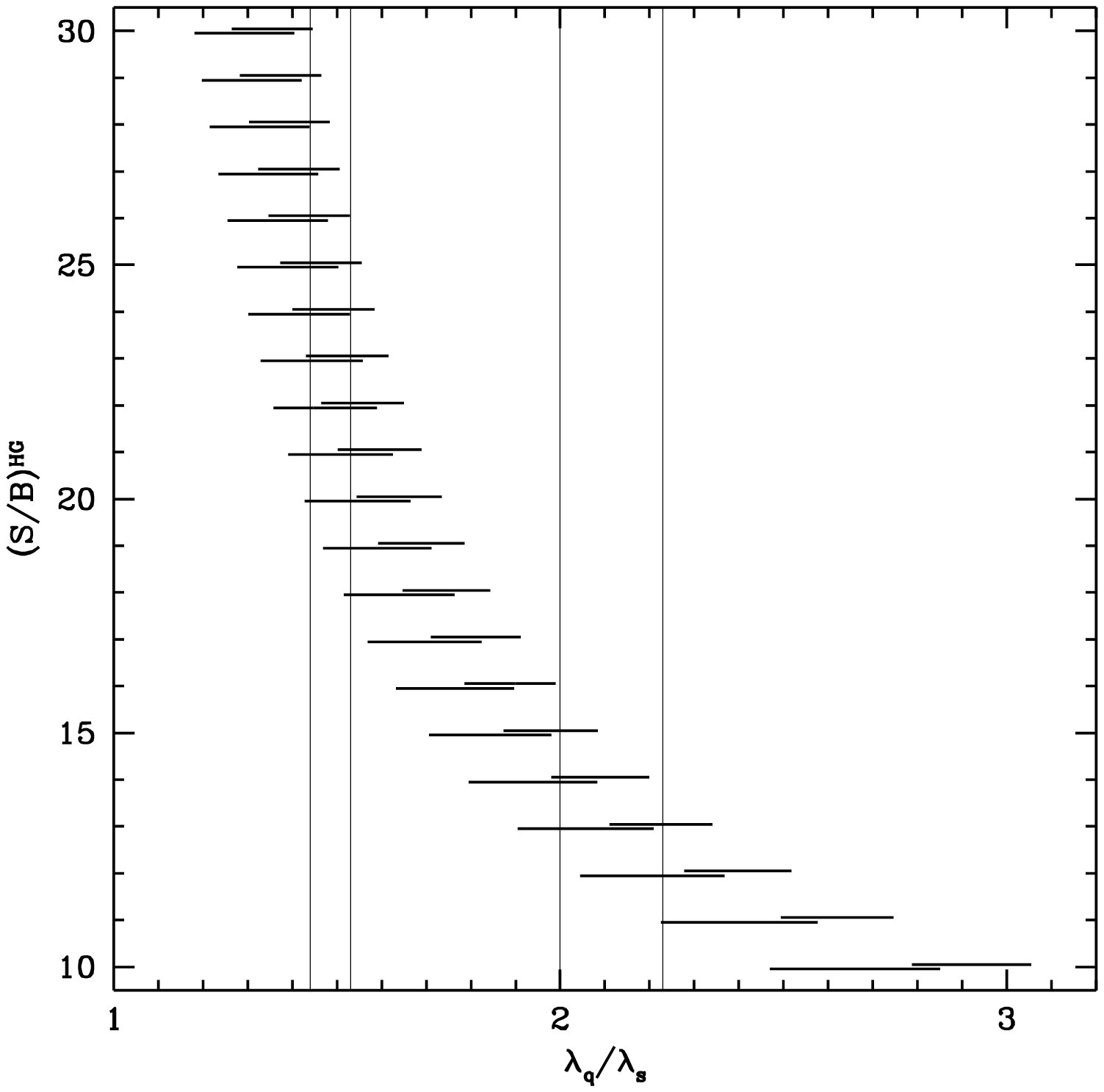}}
\vspace*{-2cm}
\centerline{\bf FIGURE 1}
\end{figure}
\newpage
\begin{figure}[p]
\centerline{\hspace{0.2cm}\psfig{figure=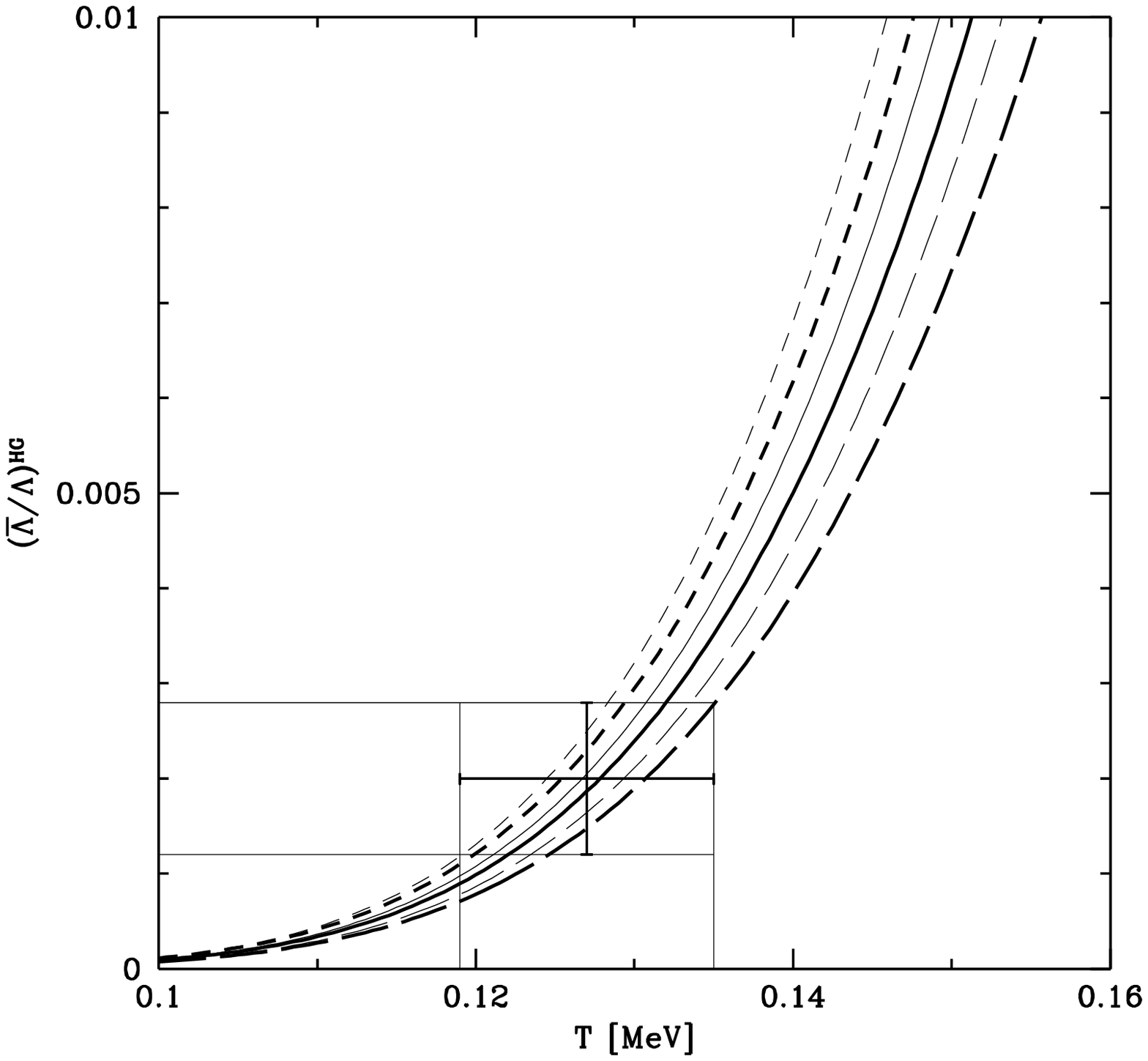}}
\vspace*{-2cm}
\centerline{\bf FIGURE 2}
\end{figure}
\newpage
\begin{figure}[p]
\centerline{\hspace{0.2cm}\psfig{figure=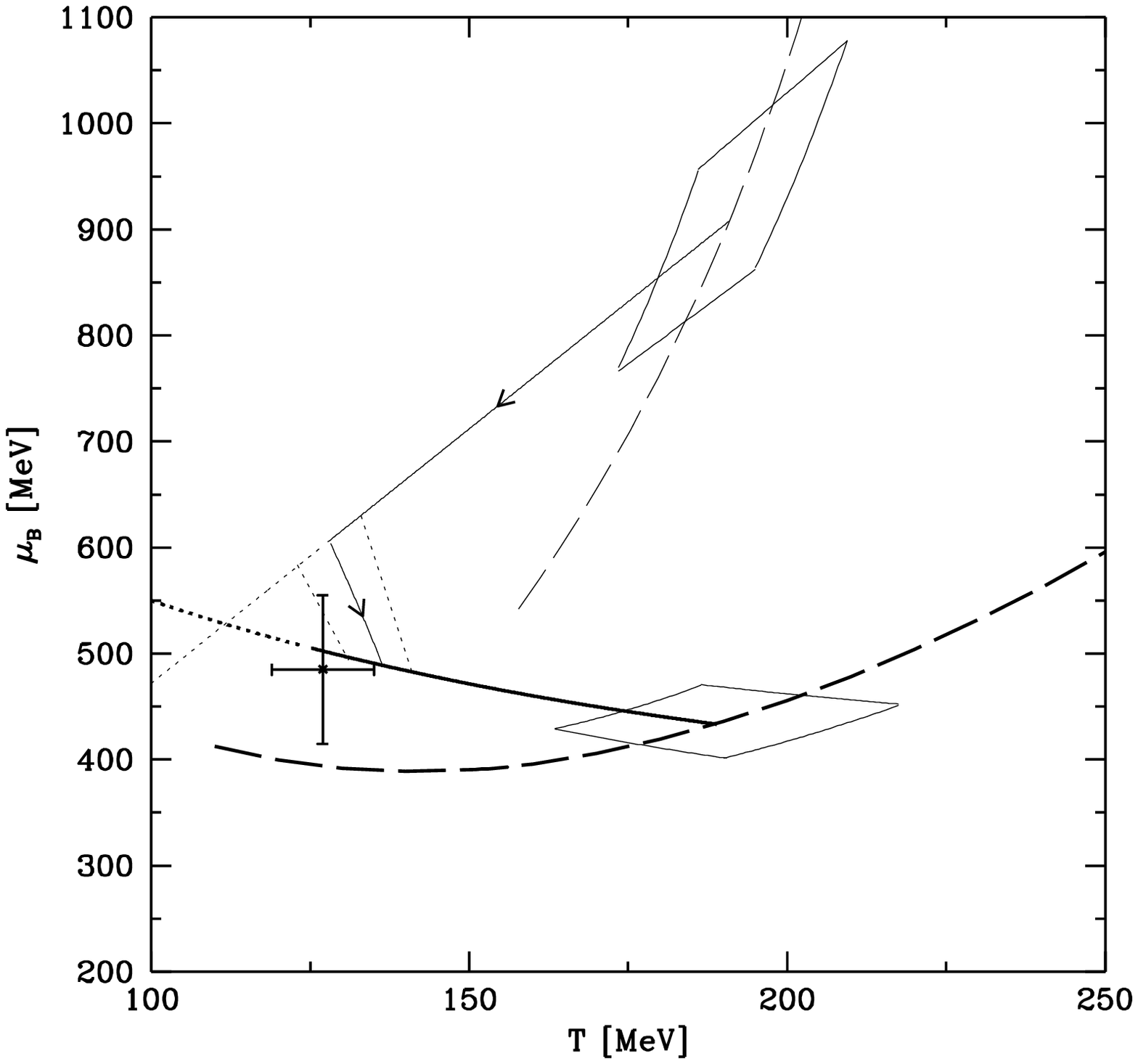}}
\vspace*{-2cm}
\centerline{\bf FIGURE 3}
\end{figure}
\end{document}